                                                                                                                                                                                                                                                                    \documentclass{amsart}
\pdfoutput=1
\usepackage{mathtools}
\usepackage{graphicx}
\usepackage{amsmath}
\usepackage{amsfonts}
\usepackage{verbatim}
\usepackage{verbatim}
\usepackage[framed,autolinebreaks,useliterate]{mcode}

\usepackage{color}
\usepackage[margin=1.5in]{geometry}

\newcommand{\half}{\frac{1}{2}}

\newcommand{\eqb}[1]{\begin{equation}\label{#1}}
\newcommand{\eqe}{\end{equation}}

\newcommand{\eqn}[2]{\begin{equation}\label{#1}#2\end{equation}}

\newcommand{\st}{\hbox{ \,\,subject to\,\, }}
\newcommand{\matb}{\left( \begin{matrix*}[r] }
\newcommand{\mate}{\end{matrix*}\right)}

\newcommand{\reals}{\mathbb{R}}

\newcommand{\mcA}{\mathcal{A}}

\DeclareMathOperator*{\minimize}{minimize\quad}

\theoremstyle{definition}

\begin{document}

\sloppy 

\bibliographystyle{ieeetr} 
\title{ {FASTA}:  A Generalized Implementation of Forward-Backward Splitting}
\author{Tom Goldstein}
\maketitle
\section{What is FASTA?}
FASTA (Fast Adaptive Shrinkage/Thresholding Algorithm) is an efficient, easy-to-use implementation of the Forward-Backward Splitting (FBS) method for minimizing compound objective functions.  FASTA targets problems of the form
 \eqb{general}
\minimize f(Ax)+g(x),
\eqe 
where $A$ is a linear operator, $f$ is a differentiable function, and $g$ is a ``simple'' function for which we can evaluate the proximal operator.  
Consider for example the $\ell_1$-penalized least squares problem
  \eqb{spls}
 \minimize  \mu |x| + \half \| Ax-b \|^2
\eqe
where $|\cdot|$ denotes the $\ell_1$ norm,  $\|\cdot\|$ denotes the $\ell_2$ norm, $A$ is a matrix, $b$ is a vector, and $\mu$ is a scalar parameter.  This problem is of the form \eqref{general} with $g(z) = \mu |z|,$ and $f(z) = \half\|z-b\|^2.$  
More generally, any problem of the form \eqref{general} can be solved by FASTA, provided the user can provide function handles to $f,$ $g,$ $A$ and $A^T.$  

The solver FASTA contains numerous enhancements of FBS to improve convergence speed and usability.  These include adaptive stepsize choice, acceleration (i.e., of the type used by the solver FISTA), backtracking line search, and numerous automated stopping conditions, and many other improvements reviews in the article {\em A Field Guide to Forward-Backward Splitting with a FASTA Implementation.}

\section{What does FASTA come with?}
Your download comes with several folders.  One folder is called {\tt solvers}.  This folder contains the file {\tt fasta.m,} which is a self-contained solver for {\em any} problem of the form \eqref{general}.  

 The {\tt solvers} folder also contains numerous specialized solvers, each of which solves a {\em specific} problem of the form \eqref{general}.  For example, the code {\tt fasta\_sparseLeastSquares} solves the sparse least squares problem  \eqref{spls}, and {\tt test\_sparseLogistic} solves $\ell_1$ penalizes logistic regression problems.  Each of these specialized solvers depends on the file {\tt fasta};  they simply cook up a specific $f,$ $g,$ and $A$ corresponding to a specific problem, and hand them off to {\tt fasta.} 

 The top-level folder contains test scripts that demonstrate how to use each solver. For example, the script {\tt test\_sparseLeastSquares} builds a random instance of a sparse regression problem and solves it using {\tt fasta\_sparseLeastSquares}.  Each of these scripts requires no setup by the user.  Simply run them from the command line.

\section{How to Install FASTA}
After you download the code, simply add the {\tt solvers} folder to your path and you're ready to go.  

Technically, you only need to add the single file {\tt fasta.m} to your path if you only want to use the general solver.  However, many of the test/demo scripts call specialized methods from the {\tt solvers} folder (or have other dependencies)  so it is best to add the whole {\tt solvers} folder to your current path.

\section{How to use FASTA}
Calling FASTA is easy.  To use the solver, you will need to implement the functions $f(x)$ and $g(x)$ and the linear operators $A$ and $A^T.$  You will also need a function {\tt grad($x$)} that generates the gradient of $f$ at $x$ and the function {\tt prox($x$,$t$)} representing the proximal mapping of $g$ at $x$ with stepsize $\tau.$  For many problems of interest, a specialized solver is already in the {\tt solvers} folder that does all this for you.  However, if you are using the general solver, you call {\tt fasta}  with the following command.
\vspace{-4mm}\begin{lstlisting}
solution = fasta(A, At, f, gradf, g, proxg, x0); 
\end{lstlisting}

Here's a complete worked example to demonstrate the use of {\tt fasta}.  Suppose we want to solve \eqref{spls}.  The script below builds a random test problem, and then solves the penalized least squares problem using {\tt fasta.}
\vspace{-4mm}\begin{lstlisting}
%% Build a simple (arbitrary) test problem
A = randn(5,10); % Define this matrix however you wish!
b = randn(5,1);  % Define this vector however you wish!
mu = 1;          % Define this scalar however you wish!

%% Build the ingredients for fasta
f = @(x)  0.5*norm(x-b)^2; % The smooth function, f
gradf = @(x)  x-b;         % The gradient of f
g = norm(x,1);             % The non-smooth function, g
proxg = @(x,t) sign(x).*max(abs(x)-mu*t,0);  % The proximal operator (shrinkage)
x0 = zeros(10,1);          % The initial guess

%% Call fasta to solve: minimize f(Ax)+g(x)
solution = fasta(A, At, f, gradf, g, proxg, x0, opts );
\end{lstlisting}

Note that for this particular problem, one could just use the built-in solver {\tt fasta\_sparseLeastSquares} by calling
\vspace{-4mm}\begin{lstlisting}
fasta_sparseLeastSquares(A,A',b,mu,x0, opts);
\end{lstlisting}
rather than using the general solver.  However, the above example demonstrates how one could build a custom solver using {\tt fasta} in the event that a specialized solver were not already available.  

\section{Slightly More Advanced usage}

A more advanced call to {\tt fasta} would look like this:
\vspace{-4mm}\begin{lstlisting}
[solution, outs] = fasta(A, At, f, gradf, g, proxg, x0, opts);
\end{lstlisting}
This method call looks a lot like what we've already seen, but with two key differences.  First, we added the argument {\tt opts}, which is a struct of options that control the behavior of {\tt fasta.}  Second, we captured the second return value {\tt outs}, which is a struct of convergence information.  We describe each of these structs below.

    By setting fields in the struct {\tt opts}, the user can control the behavior of {\tt fasta.} Several of the most useful fields are described here: \vspace{2mm}

        \hspace{-.7cm}\begin{tabular}{p{2.5cm}p{11.5cm}}
    \vspace{-3mm} \flushright {\tt opts.verbose} & Controls how much text output appears in the console. Set {\tt opts.verbose=1} for some output, and {\tt opts.verbose=2} to print convergence information after every iteration.\\
   \vspace{-3mm}  \flushright  {\tt opts.tol} & The stopping tolerance of the method. The default value {\tt tol=1e-3} works well for most problems.  However you may choose a smaller value to achieve more precision, or a larger value to achieve shorter runtime.\\
    \vspace{-3mm}  \flushright  {\tt opts.maxIters} & The maximum number of iterations the method will perform.  The default value is 1000.\\
       \end{tabular} 
  
  \vspace{2mm}
    The struct {\tt outs} contains information that can be use used to fine-tune performance.  The most commonly used outputs are:
    
     \vspace{2mm}
      \hspace{-.7cm}\begin{tabular}{p{2.5cm}p{11.5cm}}
    \vspace{-5mm} \flushright {\tt outs.solveTime } & The runtime of the algorithm.\\
    \vspace{-5mm}  \flushright  {\tt outs.residuals} & A vector containing the residuals at each iteration.  The residual is a derivative (or more generally sub-gradient) of the objective function, and should be nearly zero at a good approximate minimizer.  \\
    \vspace{-3mm}   \flushright  {\tt opts.maxIters} & The maximum number of iterations the method will perform.  The default value is 1000.\\
       \end{tabular}

\section{Specialized Solvers}

\paragraph{\textbf{Lasso Regression}}
The Lasso regression is defined as follows:
$$
 \minimize   \half \| Ax-b \|^2
 \st \|x\|_1\le \lambda.
$$
This problem is solved by calling 
\vspace{-4mm}\begin{lstlisting}
solution = fasta_lasso( A, At, b, lambda, x0);
\end{lstlisting}
where {\tt At} is the transpose of {\tt A}, {\tt lambda} is the regularization parameter, and {\tt x0} is an initial guess (usually an appropriately sized vector of zeros).

\paragraph{\textbf{{$\boldsymbol \ell_1$}-Penalized Least Squares} }
The sparse least squares (or basis pursuit denoising) problem is
$$
 \minimize  \mu \|x\|_1 + \half \| Ax-b \|^2.
$$
This problem is solved by the command
\vspace{-4mm}\begin{lstlisting}
solution = fasta_sparseLeastSquares(A, At, b, mu, x0);
\end{lstlisting}

\paragraph{\textbf{{$\boldsymbol \ell_1$}-Penalized Logistic Regression}} 
When the vector~$b\in\{0,1\}^M$ contains binary-valued entries one is interested in solving the sparse logistic regression problem
$$
 \minimize  \mu \|x\|_1 + \mathrm{logit}(Ax,b);
$$
with the logit penalty function defined as 
$$
\mathrm{logit}(z,b) =  \sum_{i=1}^{M} \log(e^{z_i}+1)  - b_iz_i. \notag
$$
This problem is solved using the following command:
\vspace{-4mm}\begin{lstlisting}
solution = fasta_sparseLogistic(A, At, b, mu, x0);
\end{lstlisting}

\paragraph{\textbf{Low-Rank (1-bit) Matrix Completion}}
FASTA can solve the matrix completion problem 
$$\minimize \mu \| X \|_* +  \mathrm{logit}(X,Y), $$
where $ \| X \|_*$ is the low-rank inducing nuclear norm of the matrix $X$ and $ \mathrm{logit}$ is the logistic loss function.  This is done with the command
\vspace{-4mm}\begin{lstlisting}
solution = fasta_logisticMatrixCompletion(B, mu);
\end{lstlisting}

\paragraph{\textbf{Phase Retrieval}}
The PhaseLift algorithm solves phase retrieval problems of the form
     $$   \minimize\!\!\!\!\! \|X\|_*  \st  \!\! \mcA(X) = b, \, X\succeq 0.$$
In the case where the measurement vector $b$ is contaminated by additive noise, we choose the  $\ell_2$-norm penalty model
        \eqn{smoothRankMin}{ 
          \minimize \!\!\! \mu\|X\|_* +\|\mcA(X) - b\|^2 \, \st  X\succeq 0, \notag
          }
which can be solved using FBS.  The solution to this problem is found by calling
\vspace{-4mm}\begin{lstlisting}
solution = fasta_phaselift(A, b, mu, X0);
\end{lstlisting}

\paragraph{\textbf{Democratic Representations}}
Given a signal $b\in \reals^M$, a low-dynamic range representation can be found by choosing a suitable matrix $A\in \reals^{M\times N}$ with $M<N$, and by solving 
  $$
     \minimize \mu \|x\|_\infty +  \half \| Ax-b \|^2.
  $$
This problem is solved by the command
\vspace{-4mm}\begin{lstlisting}
solution = fasta_democratic(A, At, b, mu, x0);
\end{lstlisting}

\paragraph{\textbf{Total Variation Denoising}}
Given a noisy image $f,$ we can find a denoised image by solving 
$$\minimize \mu |\nabla x|+\half \|x-f\|^2$$
where $|\nabla x|$ denotes the total-variation of $x.$  Denoising if performed by the command
\vspace{-4mm}\begin{lstlisting}
solution = fasta_totalVariation(f, mu);
\end{lstlisting}\vspace{-4mm}
Note:  this solver works on ``images'' of dimension 1 or higher.

\appendix 
\section{Complete list of options}
\hspace{1cm}\begin{minipage}{\linewidth}
  \begin{itemize}
    \item[\tt opts.verbose] Controls how much text output appears in the console. Set {\tt opts.verbose=1} for some output, and {\tt opts.verbose=2} to print convergence information after every iteration.
    \item[\tt opts.tol] The stopping tolerance of the method. The default value {\tt tol=1e-3} works well for most problems.  However you may choose a smaller value to achieve more precision, or a larger value to achieve shorter runtime.
     \item[\tt opts.maxIters] The maximum number of iterations the method will perform.  The default value is 1000.
     \item[opts.recordObjective] If {\tt opts.recordObjective=true,} then the method will evaluate the objective function $f(x_k)+g(x_k)$ at every iteration and store the results in {\tt outs.objective}.  Computing the objective takes time, and so turning this option on may slow down computation for some problems.  The default is {\tt opts.recordObjective=false,}
      \item[opts.recordIterates]  If {\tt opts.iterates=true}, then every iterate of the method is stored and returned in the cell array {\tt outs.iterates}.  This option is turned off by default.  Turning it on may dramatically increase memory requirements. 
      \item[opts.adaptive]  Determines whether adaptive stepsizes are used.  By default {\tt opts.adaptive=true.}
      \item[opts.accelerate]  Determines whether to use the accelerated method FISTA. By default this is turned off, but can be turned on by setting {\tt opts.accelerate=true}.  If this option is turned on, then the user may assign a boolean value to {\tt opts.restart} to determine whether to use a ``restart'' rule (default behavior uses restart).
      \item[opts.function]  The user may supply a function that takes a single argument.  On every iteration, the value of  {\tt opts.function($x_k$)} is computed and stored in the cell array  {\tt outs.funcVals.}
      \item[opts.backtrack]  Determines whether backtracking is use to guarantee stability.  If this option is set to {\tt false},  then the user should either set the stepsize manually in {\tt opts.tau}, or else supply a Lipschitz constant for $\nabla f$ in {\tt opts.L}.   By default {\tt opts.backtrack=true,} and there is frequently no benefit in turning this option off.
      \item[opts.stopRule]  A string that determines which stopping condition is used.  Choose a value from {\tt \{ratioResidual, normalizedResidual, hybridResidual\}}.  A hybrid residual strategy is used by default.
      \item[opts.stopNow] The user may implement a custom stopping rule.  At each iteration $k$, the function {\tt opts.stopNow($x_k$,k,residual,normalizedResidual,maximumResidual,opts)} is evaluated. Iteration stops when the returned value is {\tt true}.  When this {\tt opts.stopNow} is defined, this function overrides the built-in stopping rules.
     \item[opts.stringHeader] This string is appended to the front of all text output when {\tt opts.verbose=true}.  This option allows the user can add custom labels to text that is printed to the console.
    \end{itemize} 
     \end{minipage}

\newpage
\section{Complete list of Outputs}
    The struct {\tt outs}, contains convergence information that can be use used to fine-tuning convergence.  The outputs are:
    
       \hspace{1cm}\begin{minipage}{\linewidth}
    \begin{itemize}
    \item[\tt outs.solveTime] The runtime of the algorithm.
    \item[\tt outs.residuals]  A vector containing the residuals at each iteration.
    \item[\tt outs.stepsizes]   A vector containing the stepsize used at each iteration.
    \item[\tt outs.normalizedResiduals]  The normalized residuals at each iteration.
    \item[\tt outs.objective]  The objective function evaluated at each iterate.  This is not recorded by default.  Set {\tt opts.recordObjective=true}  to use this option.
    \item[\tt outs.funcValues]  Stores the values of  {\tt opts.function($x_k$)} for each iterate $x_k.$ If the user did not supply a value for  {\tt opts.function,} then this will be a vector of zeros.
    \item[\tt outs.backtracks] The number of times backtracking was activated.
    \item[\tt outs.L]   The estimated Lipschitz constant for $\nabla f.$
    \item[\tt outs.initialStepsize]  The initial stepsize used for the first iteration.
    \item[\tt outs.iterationCount]  The total number of iterations computed before termination.
    \item[\tt outs.iterates]     If {\tt opts.recordIterates=true,} then this field is a cell array containing every iterate of the method.
        \end{itemize} 
     \end{minipage}

\bibliography{/Users/Tom/Documents/latexDocs/bib/tom_bibdesk}

\end{document}